\documentclass[]{aa}

\usepackage{graphicx,txfonts,lscape}
\bibpunct{(}{)}{;}{a}{}{,} 

\begin{document}

   \title{Exploring the substellar population in the Hyades open cluster}


   \author{A.\ P\'erez-Garrido \inst{1}
        \and
        N.\ Lodieu \inst{2,3}
        \and
        R.\ Rebolo \inst{2,3,4}
        \and
        P.\ Chinchilla\inst{2,3}
        }

   \institute{Dpto.\ F\'\i sica Aplicada, Universidad Polit\'ecnica de Cartagena, E-30202  Cartagena, Murcia, Spain\\
       \email{antonio.perez@upct.es}
       \and
       Instituto de Astrof\'isica de Canarias (IAC), Calle V\'ia L\'actea s/n, E-38200 La Laguna, Tenerife, Spain
       \and
       Departamento de Astrof\'isica, Universidad de La Laguna (ULL), E-38206 La Laguna, Tenerife, Spain
       \and
       Consejo Superior de Investigaciones Cient\'ificas, CSIC, Spain
             }

   \date{Received \today{}; accepted (date)}

 
  \abstract
   {}
   {Our aim is to identify substellar members of the nearby  Hyades open star cluster to determine the photometric and spectroscopic properties of brown dwarfs at moderately old ages  and extend our  knowledge  of the substellar mass function of the cluster.
   }
   {We cross-matched the 2MASS and WISE public catalogues and measured proper motions to  identify low-mass 
stars and brown dwarf member  candidates in an area of a radius of ten degrees around the central region of the 
Hyades cluster.  We employed astrometric and photometric criteria, Gaia data, and a maximum likelihood method developed by our group to estimate distances. We selected 36 objects that are candidate Hyades members, 21 of which have not been reported previously.
  
   }
   {
We have identified 21 new Hyades member candidates that are placed at the lower end of the main sequence.
 The photometry of 9 candidates places them in the substellar regime, and 2 are at the L/T transition.
We also recovered a number of L dwarfs from earlier surveys.   Finally, we calculated the mass function for the low-mass population of the cluster and found that the Hyades cluster might have lost 60\%\ to 80\% of its substellar members.  
 }
   {}  

  \authorrunning{P\'erez Garrido et al$.$}
  \titlerunning{Low-mass objects of the Hyades}

   \maketitle
%

%
%
\section{Introduction}
\label{Hyades:intro}

Brown dwarfs, which are objects with masses below 0.075 $M_{\odot}$, play an important role in the understanding of processes that are related to 
star or substellar formation and evolution.  The temperatures at their cores are not high enough to allow for hydrogen fusion \citep{burrows93,chabrier97}. The lack of sustained hydrogen burning causes these objects to change physical properties with time.
Older star clusters ($>$\,500\,Myr) like the Hyades (Melotte 25, $\alpha_{2000}$\,=\,04$^{\rm h}$26$^{\rm m}$54$^{\rm s}$, 
$\delta_{2000}$\,=\,$+$15$\hbox{$^\circ$}$52$\hbox{$^\prime$}$)  are excellent testbeds for
studying brown dwarfs with known age and metallicity. The general properties of the Hyades cluster can be found in the literature (see e.g. \cite{perezgarrido17}).
The age of the Hyades is 625$\pm$50 Myr based on the comparison 
of the observed cluster sequence with model isochrones. Recently, \cite{martin18} estimated an age of 650$\pm70$ Myr using the lithium  depletion boundary. \cite{lodieu18} calculated upper limits for the age of the Hyades of 775 and 950 Myr from the
luminosity and T$_{\rm eff}$ of a confirmed L5 dwarf Hyades member.
Even so, a wider age range (500--800 Myr) cannot be discarded 
\citep{mermilliod81,eggen98a,brandt05}.    A recent study employing Gaia photometry yielded an age of 794 Myr \citep{gaia18b}.
The metallicity of the Hyades high-mass stars appears slightly supersolar, with values between 0.127 $\pm$0.022 and 0.14 $\pm$0.1 \citep{boesgaard90,cayrel97,grenon00}, although more recently, \cite{gebran10} suggested a mean metallicity close 
to solar ([Fe/H] = 0.05 $\pm$0.05).  Photometry from the Gaia mission also gives a metallicity of [Fe/H]\,$\sim$\,0.13 dex 
\citep{gaia18b}.

 \citet{hogan08} reported 12 L dwarf candidates in the Hyades cluster. \citet{casewell14a} and \citet{lodieu14b}  presented spectroscopic follow-up that confirmed the cool nature of most of them, with a lithium depletion boundary around spectral type L3-L4 \citep{martin18}. 
\cite{bouvier08a} discovered the first two T-type dwarfs in the Hyades cluster by means of
low-resolution infrared spectra and  claimed that $\sim$15 brown dwarfs 
could exist in the present-day Hyades cluster.
In addition to 2MASSI\,023301.55$+$2470406 \citep{cruz07}, which was proposed as an L0 member of the Hyades \citep{goldman13}, \cite{schneider17} very recently also found and spectroscopically confirmed two L dwarfs placed in the region of the Hyades cluster.
A few other known L/T dwarfs listed with spectra and proper motions in the compendium of
ultracool dwarfs\footnote{see http://spider.ipac.caltech.edu/staff/davy/ARCHIVE/index.shtml} 
could be associated with the Hyades moving group \citep{bannister07,gagne15c} and therefore could share the age of the Hyades members.

In this paper, we present new ultracool member candidates of the Hyades cluster. One of them, 2MASS\,J0418$+$2131,  was  studied spectroscopically  by our group and was confirmed as a new substellar Hyades member with the detection of lithium in absorption and H$\alpha$ emission due to chromospheric activity \citep{perezgarrido17,lodieu18}.


%
%
\section{New very low mass proper motion members of the Hyades}
\label{Hyades:new_memb}
\subsection{Catalogue cross-match and candidate selection}
\label{Hyades:new_memb_Xmatch}

We cross-matched the Two Micron All Sky Survey  point 
source catalogue \citep[2MASS;][]{cutri03,skrutskie06} and the mid-infrared catalogue built upon the Wide-field Infrared Survey Explorer 
mission \citep[WISE;][]{wright10} to uncover new low-mass stars and substellar members in the Hyades cluster. In particular, we searched in a circular region with radius of 10 degrees.  Details of the matching procedure are reported in \cite{perezgarrido17}. With this method we compiled a preliminary list of moving objects in the Hyades region.
In order to estimate the error in our proper motion calculations, we 
checked the difference in coordinates of quasars in the region. We used the Million Quasars Catalog (MILLIQUAS), version 5.2 \citep{flesch17}, and found about 100 quasars with data in 2MASS and WISE.  We averaged the absolute value of the difference in coordinates between both catalogues and obtained a positional uncertainty of 222 mas. As the WISE-2MASS epoch separation is about 11.5 years,
our proper motion estimations have an error of 19.3 mas/yr.

\begin{figure}
\resizebox{\hsize}{!}{\includegraphics{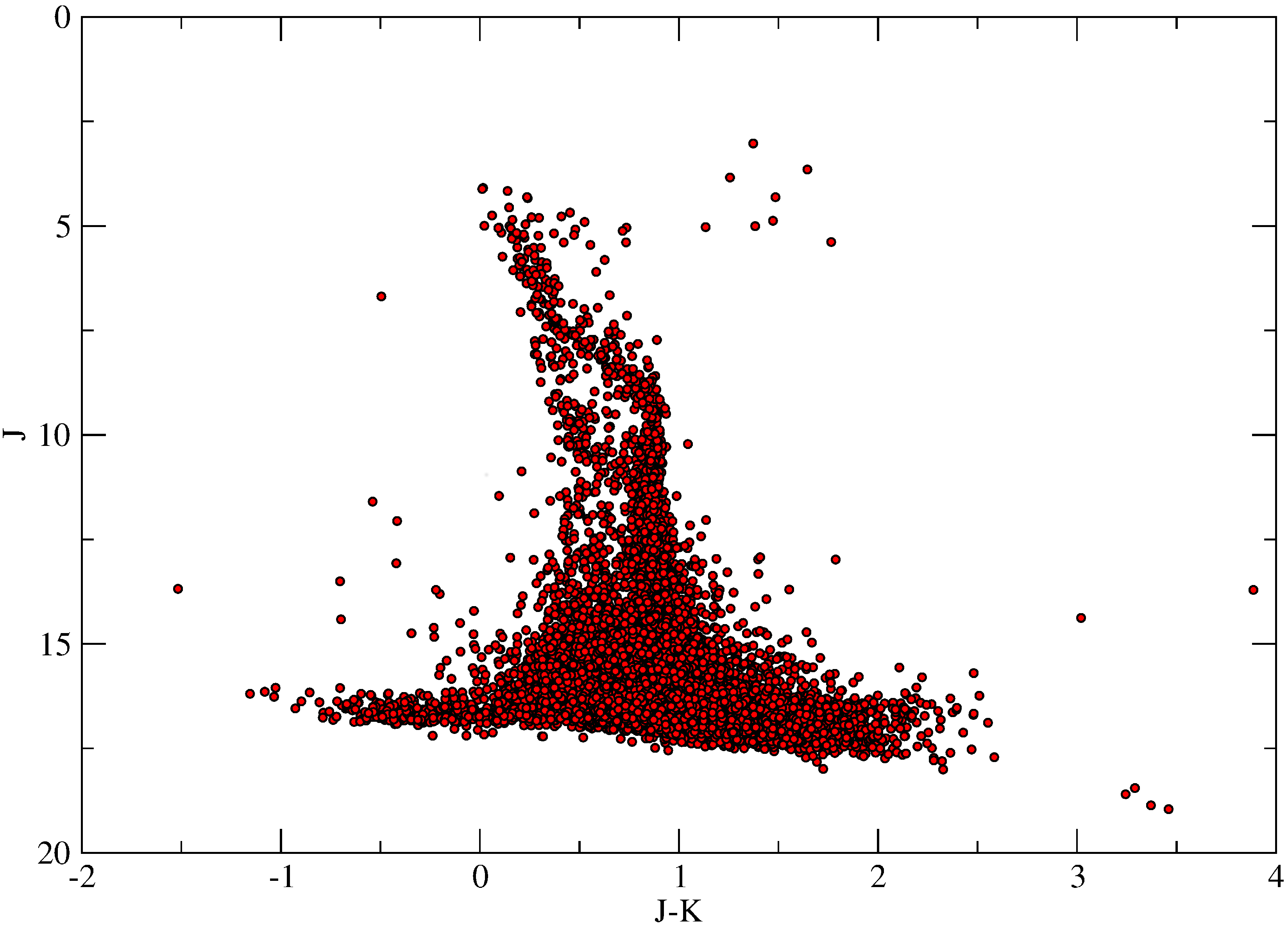}}
\caption{($J-K$,$J$) colour-magnitude diagram for objects found in our cross-match with a proper motion consistent with Hyades cluster membership. 
}
\label{fig_Hyades:total_J_J-K}
\end{figure}

\begin{figure}
\resizebox{\hsize}{!}{\includegraphics{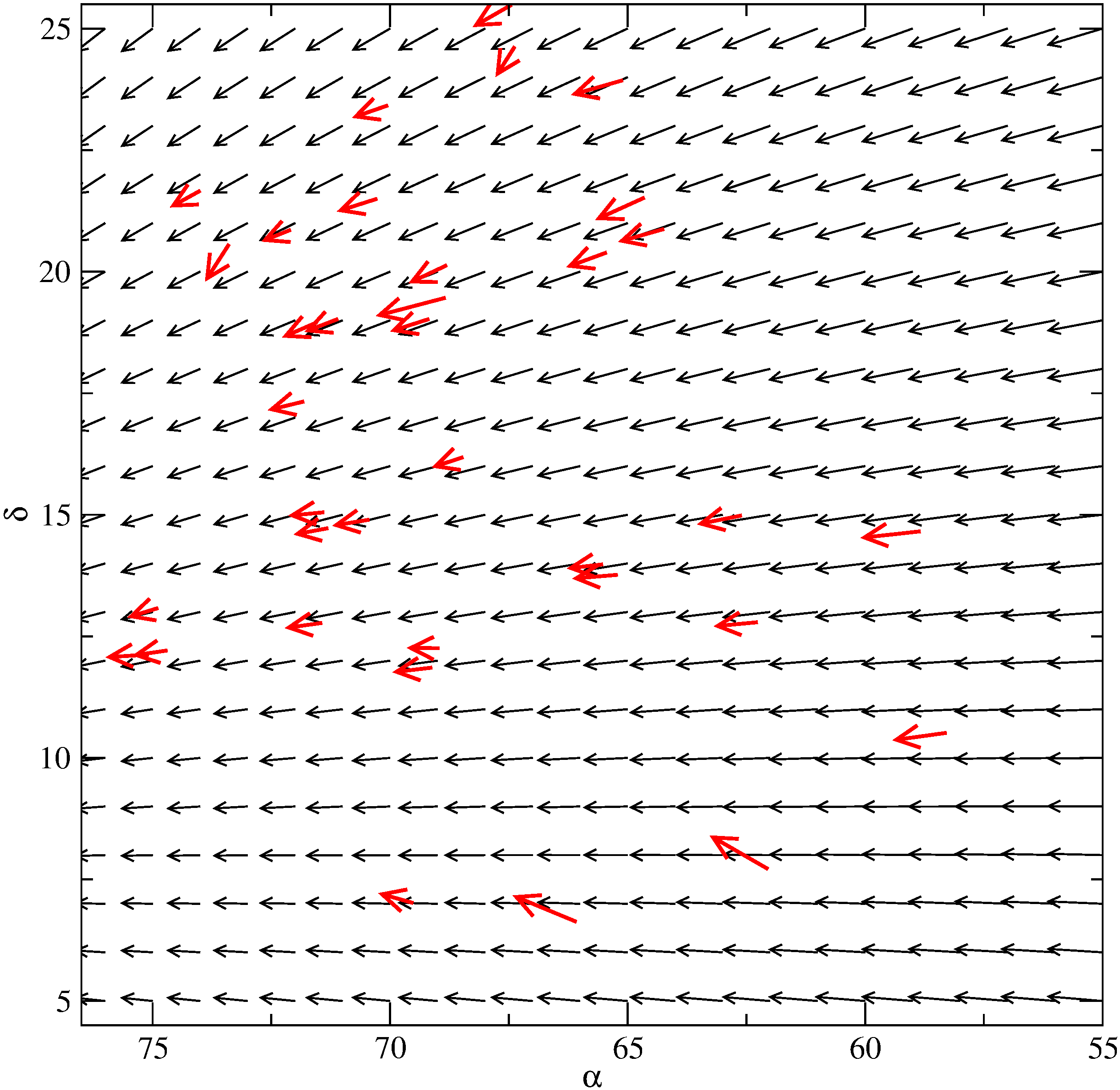}}
\caption{Angular  velocity map of the Hyades cluster. Black arrows correspond to the average proper motion obtained from the convergent point method,
 and the red arrows correspond to the proper motion from objects in our final list of candidates.
}
\label{fig_Hyades:PM}
\end{figure}

In a subsequent step, we selected objects by applying the cluster convergent point (CP) method as described in \cite{hogan08}. For the sake of completeness, we briefly explain the method and include the main equations. The CP is a single point where all members of a cluster seem to be moving and whose coordinates are
 $\alpha_0=6^{\rm h}29.48^{\rm m}$, $\delta_0=6^{\circ}53^\prime.4$ for the Hyades cluster \citep{madsen02}. 
For each source, we calculate $\theta$: the  angle between the 
north line and the line passing through the object position and the CP.  Firstly, we need to know the angular distance, $D$, between the CP and each object:
\begin{equation}
\cos D =\sin\delta\sin \delta_0$+$\cos\delta\cos\delta_0\cos \left(\alpha-\alpha_0 \right),
\end{equation}
where $\alpha$ and $\delta$ are the RA and Dec.\ of each object, respectively. The angle $\theta$ is calculated using the equation
\begin{equation}
\cos\theta=\frac{\sin\delta_0-\sin\delta\cos D}{\cos\delta\sin D}.
\label{eq_Hyades:theta}
\end{equation}
In a first step, we discard objects with $\left|\theta_{\rm PM}-\theta\right|>\theta_{max}^{\circ}$, where $\theta_{\rm PM}$ is the angle between the north line and the proper motion vector. The Hyades members have a proper motion $\sim100$ mas/yr and the  calculated proper motion accuracy is of 19.3 mas/yr, thus, the observed angles have an error of $\tan^{-1} (19.3/100)=11^\circ$. In order to select objects within 3$\sigma,$ we choose an angle $\theta_{max}=$33$^\circ$.
This guarantees that we keep in our list objects that are moving towards the CP. A second criterion takes into account the total proper motion $\mu$. The theory of moving clusters allows estimating the distance $d$ of a member with the expression
\begin{equation}
d=\frac{46.7\sin D}{4.74\mu},
\label{eq_Hyades:d}
\end{equation}
thus, we can assess whether the proper motion modulus $\mu$ is consistent with the expected cluster proper motion modulus. In Eq.\ \ref{eq_Hyades:d},
$\mu$ is measured in arcsec/yr and $d$ in pc. 
Then, as a second step, all objects with $\left| d-d_{\rm Hyades}\right| > l_{max}$ are removed from our list ($d_{\rm Hyades}=46.3$ pc). 
$l_{max}=r+\Delta$, where $r=10.5$ pc is the cluster tidal radius and $\Delta$ is to account for the proper motion estimation uncertainty. As we have an error of 19.3\%, our $\Delta=8.5$ pc and then $l_{max}=19.4$ pc. 
The rigorous justification of these criteria are described in \cite{hogan08} and references therein. This selection method returned $\sim$14000 objects, which we plot in Fig.\ \ref{fig_Hyades:total_J_J-K}.  The cluster main sequence is clearly discernible in the ($J-K$,$J$) colour-magnitude diagram plotted in Fig.\ \ref{fig_Hyades:total_J_J-K}.  However, this astrometric selection contains many contaminants that must be removed, especially at magnitudes fainter than $J\sim15$. We follow a set of procedures to correct this list for contamination as explained below.
In Fig.\ \ref{fig_Hyades:PM} the average proper motion from $\theta$ and $\mu$ obtained from Eqs.\ \ref{eq_Hyades:theta} and \ref{eq_Hyades:d} is plotted as a function of position in a vector map diagram (black arrows). Red arrows represent the proper motion of objects that are included in our final list of Hyades candidates.

\begin{figure}
  \includegraphics[width=\linewidth, angle=0]{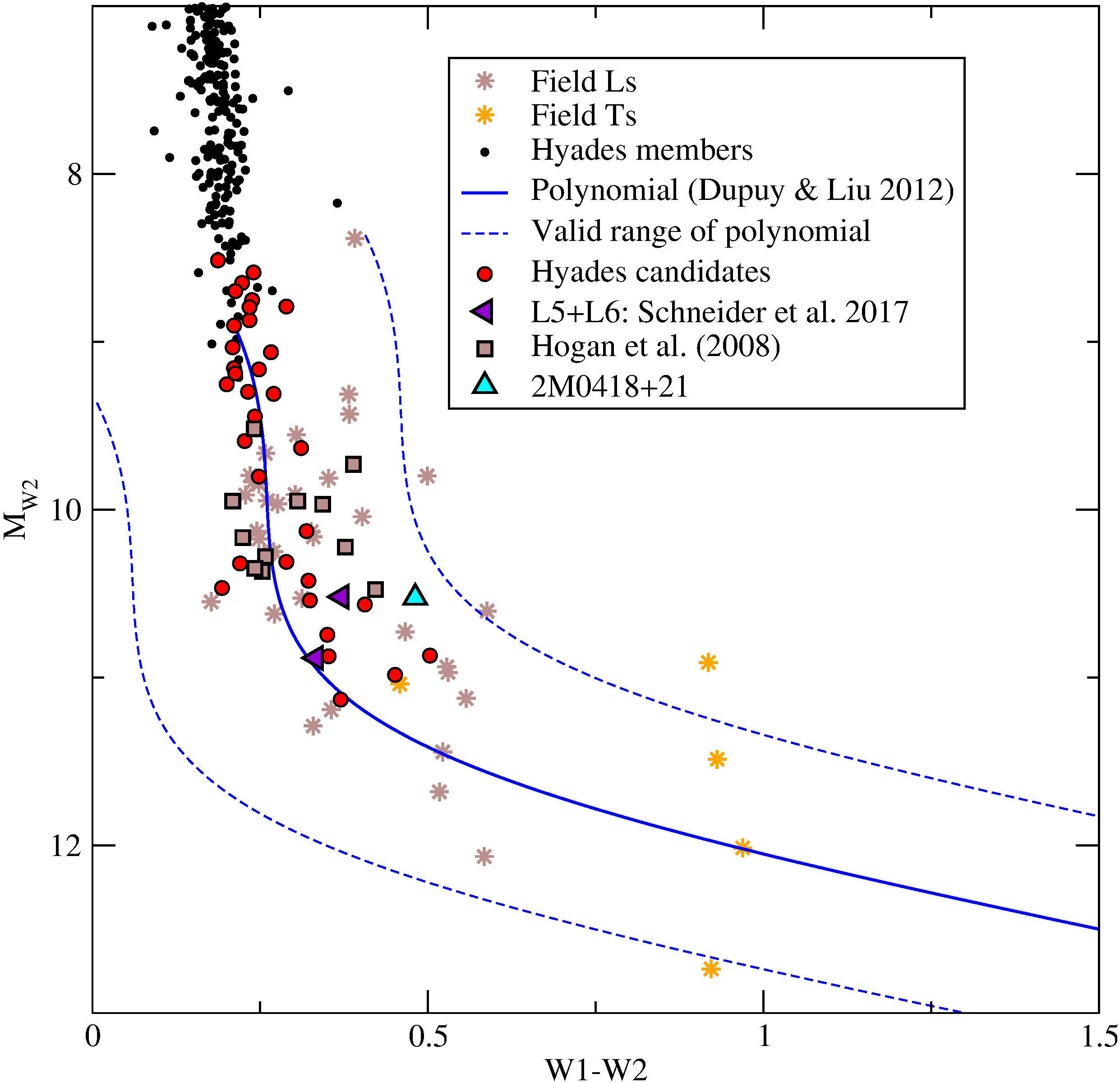}
\caption{($W1-W2$,M$_{W2}$) colour-magnitude diagram depicting known L dwarfs in the Hyades 
\citep[squares;][]{hogan08,casewell14a,lodieu14b}, \citep[left-pointing triangles;][]{schneider17}
Hyades high-mass and low-mass members 
\citep[black dots;][]{goldman13}, 
field L and T dwarfs (brown and yellow asterisks, respectively),
the mean sequence of field L and T dwarfs 
\citep[blue line;][]{dupuy12}, our confirmed L5 2M0418$+$21 \citep[light blue triangle;][]{perezgarrido17,lodieu18}, and our candidates (solid red circles). Absolute magnitudes are calculated assuming that Hyades objects and candidates are at 46.3 pc.
}
\label{fig_Hyades:MW2}
\end{figure}

To correct our astrometric sample for contaminants, we employed a photometric criterion. In a first step, using a polynomial fit extracted from Table 14 in \cite{dupuy12}, we calculated the expected absolute magnitude in $W2$ band and the $W1-W2$ colour as a function of the spectral type, $spt$, obtaining a curve defined by a set of pairs $\lbrace W2_{spt},(W1-W2)_{spt}\rbrace$. Two additional curves can be defined as some sets of pairs in the form 
$\lbrace W2_{spt}$+$0.5,(W1-W2)_{spt}-0.2\rbrace$ and $\lbrace W2_{spt}-0.5,(W1-W2)_{spt}$+$0.2\rbrace$. These three curves are shown as dashed lines in Fig. \ref{fig_Hyades:MW2}. The latter pair of curves establishes a region for good photometric candidates. Objects outside the area delimited by these curves were removed from our list. The Hyades cluster is an old cluster, but younger than field L/T. Thus when using these polynomials, calculated using field objects, we must be conservative since younger L dwarfs are usually redder in these bands.
As we are interested in the brown dwarf population, we removed objects with $J-K<1$, although we previously checked that no T-type brown dwarf (with bluer $J-K$ colour) was lost.
The absolute magnitude of T0 dwarfs is $M_J\sim 14.6,$ and the  nearest objects in our search are located at about 27 pc,
 which means that they have an expected magnitude at these distances larger than $J\sim 17$. This is too faint to be included in the 2MASS catalogue.
In a final step, a visual inspection was carried out to discard residual contaminants, mostly extended objects and double stars that are unresolved in WISE images.

After the steps mentioned above, 202 objects remained in our list. The  proper motions and near-infrared photometry of all these objects are compatible with cluster membership (Figs.\ \ref{fig_Hyades:PM} and\ \ref{fig_Hyades:MW2}). 
In this list we recovered several candidate M/L dwarfs from \cite{hogan08}, but we failed to find three of them: Hya07 and Hya10 do not have a counterpart in AllWISE, and a third object (Hya11) is located beyond 10 degrees from the centre of the cluster. 
We also recovered in our search the two substellar objects that have recently been found by \cite{schneider17}, see Table \ref{tab_Hyades:Candidates}. 
The T dwarfs found by \cite{bouvier08a} in the Hyades clusters (CFHT-Hy-20 and CFHT-Hy-21)
are too faint and are not included in the 2MASS point source catalogue. All selected objects have compatible proper motions and photometry,  but even so, our list  of candidates is still populated by a large number of contaminants.

To refine our selection, we used parallax data from the recently released Gaia DR2 \citep{gaia16,gaia18}. We selected from our list the sources with parallactic distances in the range 35.8--56.8 pc, and used as before a cluster mean distance of 46.3 pc and a tidal radius of 10.5 pc. We took into account the error in parallax to determine whether an object lay within the allowed range of distances.
The faintest objects in our list of candidates, those later than $\sim$L1 ($G\approx$ 20.5--21),  are not included in Gaia catalogue.
To complete the selection for objects without records in the Gaia database, we developed a maximum likelihood method to estimate  distances using their photometric data. These estimated distances were used to evaluate their inclusion in the final list. Let $p_c(spt)$ be the polynomial fit from \cite{dupuy12}, giving the colour $c$ as a function of the spectral type, $spt$ ($p_c(spt)$ is the difference of two polynomials) and let $c_i$ be the colour $c$ from object $i$, where $c\in \mathbb{C} =\lbrace J-H,H-K_s,K_s-W1, W1-W2 \rbrace$. Colours were chosen to sample different regions of the spectrum, therefore they can be considered as independent variables. We define the following likelihood function for each object $i$:
\begin{equation} 
\mathcal{L}_i(spt)=\prod_{c\in \mathbb{C}} 
\exp\left\lbrace -\frac{\left(c_i-p_c(spt)\right)^2} {\left(\sigma_{i;c}+\sigma_{c}\right)^2  }\right\rbrace,
\label{likelihood}
\end{equation}
where $\sigma_{i;c}$ is the photometric error in colour $c$ for object $i$ and $\sigma_c$ is the rms in the polynomial fit; see the rightmost column in Table 14 from \cite{dupuy12}. The $spt$ that maximizes Eq.\ \ref{likelihood} gives us a good estimate of the spectral type for each object in our sample. The usual procedure in maximum likelihood methods is to find the parameter set that verifies
\begin{equation}
\frac{\partial \ln \mathcal{L}_i(\alpha)}{\partial \alpha}=0.
\end{equation}
 The likelihood function, Eq.\ \ref{likelihood}, has just one parameter: the spectral type. It
is therefore straightforward to find the maximum of $\mathcal{L}_i$ by scanning the function for a fine grid of values of $spt$. 
In a further  step, we used the spectral type estimate to compute a photometric distance using the difference between the expected absolute magnitude (again by means of the polynomial fits from Dupuy et al.\ 2012) and magnitudes in bands $J$, $H$, $K_s$, $W1$ and $W2$ from catalogues. We calculated the distance as the average of the distances obtained for each band and determined the error as the dispersion in these five estimates. Then we considered as bona fide candidates objects with distances in the interval 46.3$\pm (10.5+3\epsilon_i)$ pc, where 10.5 pc is the tidal radius and $\epsilon_i$ is the error in the distance estimate for object $i$. In Table \ref{tab_Hyades:Candidates} we show the photometric data of the {candidates found in this search.  Fifteen of them have been confirmed as members of the cluster}. In this table we also include data from the Sloan Digital Sky Survey DR12 \citep[SDSS;][]{alam15,york00}.
In Table \ref{tab_Hyades:likelihoodResults} we list the astrometric data of our selected sample: the proper motions obtained in our cross-match, and for brighter objects, the proper motions measured by Gaia. In this table we also list the photometric distance obtained by the likelihood method and parallactic distance from Gaia. The last two columns correspond to the spectral type estimates for each source, together with their reported spectral types. By comparing the spectral type estimates with the real types, it is possible to assess the goodness of our likelihood method. The difference typically is 1 SpT or less, except for the latest L dwarfs.

%
%

%

%

%

\begin{figure}
  \includegraphics[width=\linewidth, angle=0]{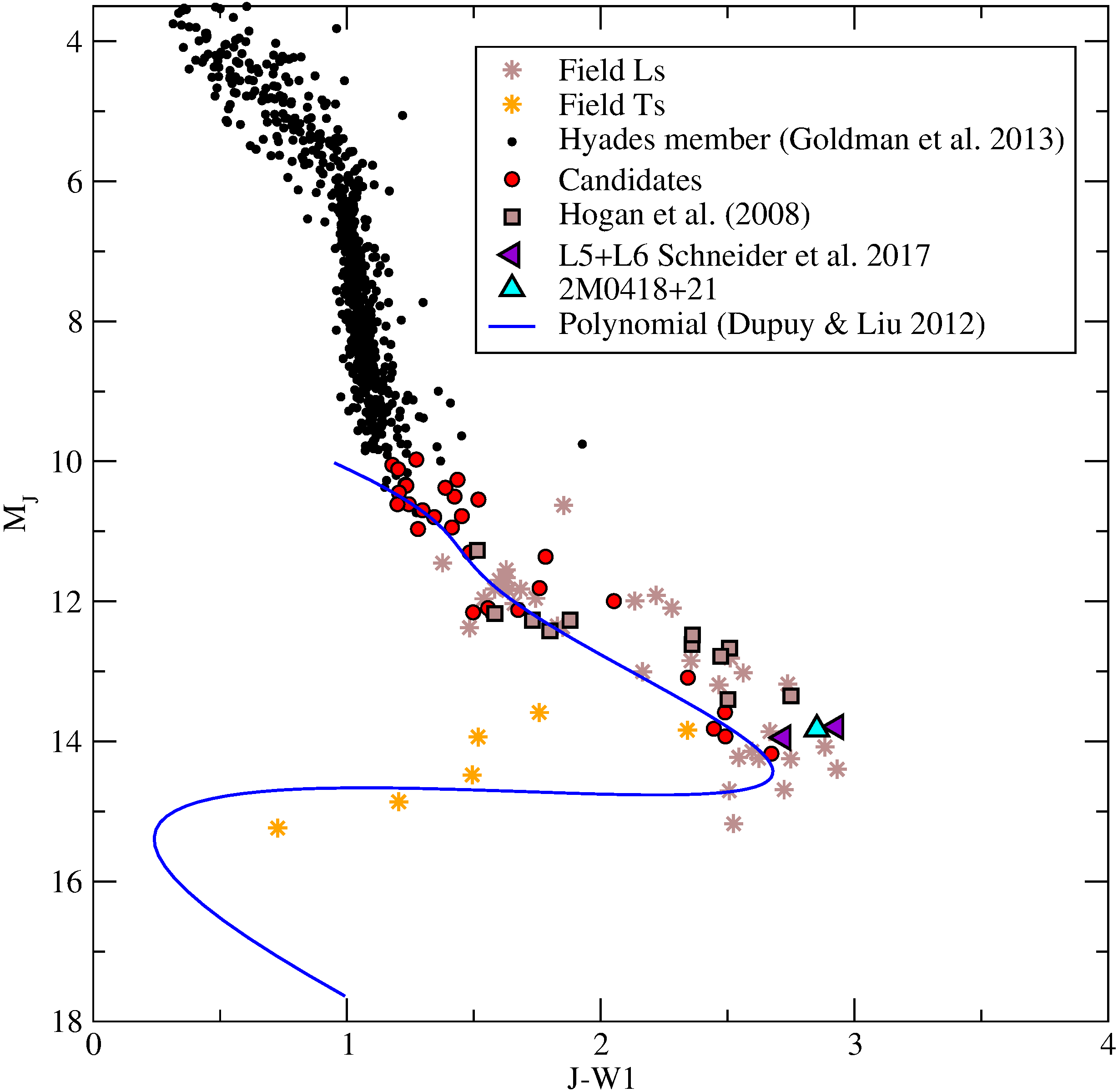}
\caption{($J-W1$,M$_{J}$) colour-magnitude diagram depicting known L dwarfs in the Hyades. Symbols  are the same as in Fig.\ 
\ref{fig_Hyades:MJ-J-W1}. 
}
\label{fig_Hyades:MJ-J-W1}
\end{figure}

\begin{figure}
  \includegraphics[width=\linewidth, angle=0]{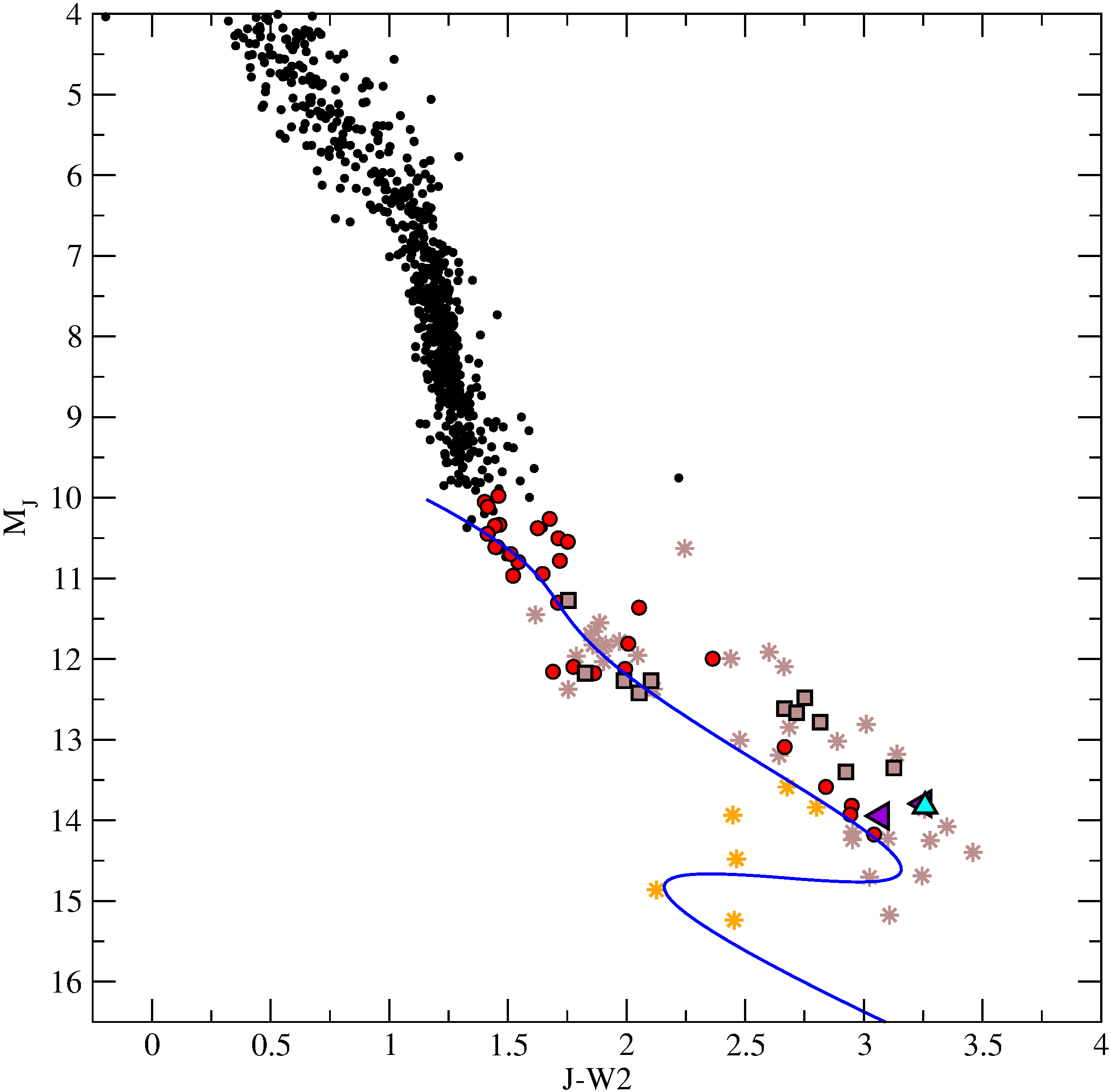}
\caption{($J-W2$,M$_{J}$) colour-magnitude diagram depicting known L dwarfs in the Hyades. Symbols are the same as in Fig.\ 
\ref{fig_Hyades:MJ-J-W1}. 
}
\label{fig_Hyades:MJ-J-W2}
\end{figure}

\section{Substellar candidates}

In our search we selected a number of objects that span from late-M to objects deep in the L dwarf region. In particular, we found eight new objects with an estimated L spectral type. 
2M0424$+$0637 has the latest estimated spectral type (L9.1), with the reddest $W1-W2=0.5$ and $i-z=2.74$ colours, which places it at about the L--T boundary. This object shows a slightly discrepant proper motion (see Table \ref{tab_Hyades:likelihoodResults}).  2M0424$+$0637 is a very faint object, close to the 2MASS limiting magnitude ($J\sim 17$); its astrometry is prone to larger error. We recomputed the proper motion using data from the SDSS and WISE catalogues and obtained $\mu_\alpha=100$ mas/yr and $\mu_\delta=4$ mas/yr, which fits the cluster motion much better. A similar object, 2M0433$+$1611, shows a slightly bluer $W1-W2=0.45$ colour and an estimated spectral type of L8. If these estimates and the Hyades membership were confirmed, they would be the latest L dwarfs of the cluster, filling the low-mass wing of the Hyades main sequence to the early T dwarfs. The mid-L dwarf region is almost depleted of candidates or members and has a much lower object density than the earliest and latest L regions, as can be appreciated in Table \ref{tab_Hyades:likelihoodResults} and in Figs.\ \ref{fig_Hyades:MJ-J-W1} ($J-W1\sim2$)  and  \ref{fig_Hyades:MJ-J-W2} ($J-W2\sim2.25$).

%
%

\begin{figure}
  \includegraphics[width=\linewidth, angle=0]{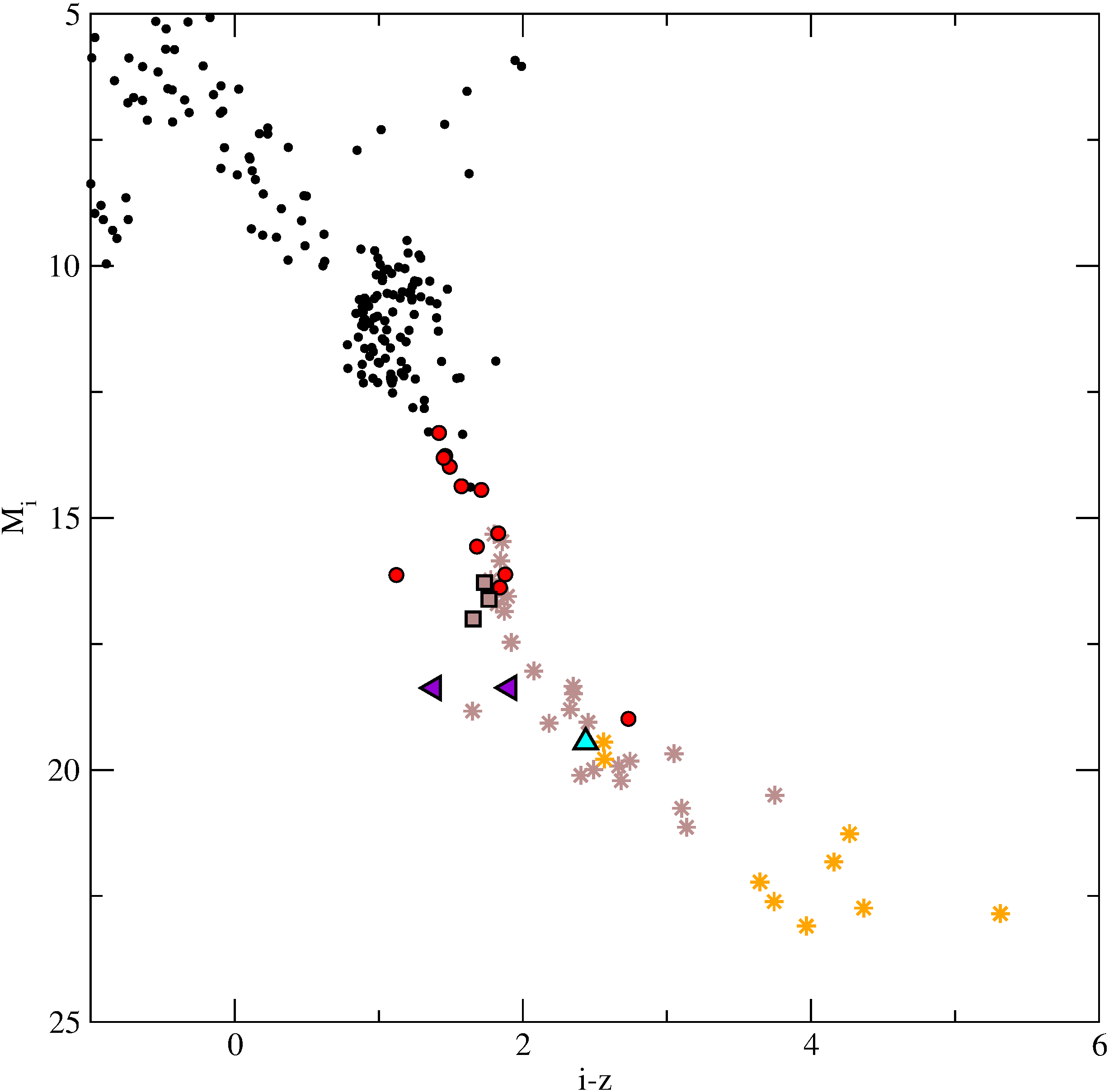}
\caption{($i-z$,M$_{i}$) colour-magnitude diagram for objects with data in the SDSS. Symbols are the same as in Fig.\ 
\ref{fig_Hyades:MJ-J-W1}. 
}
\label{fig_Hyades:Mi-i-z}
\end{figure}

\begin{figure}
  \includegraphics[width=\linewidth, angle=0]{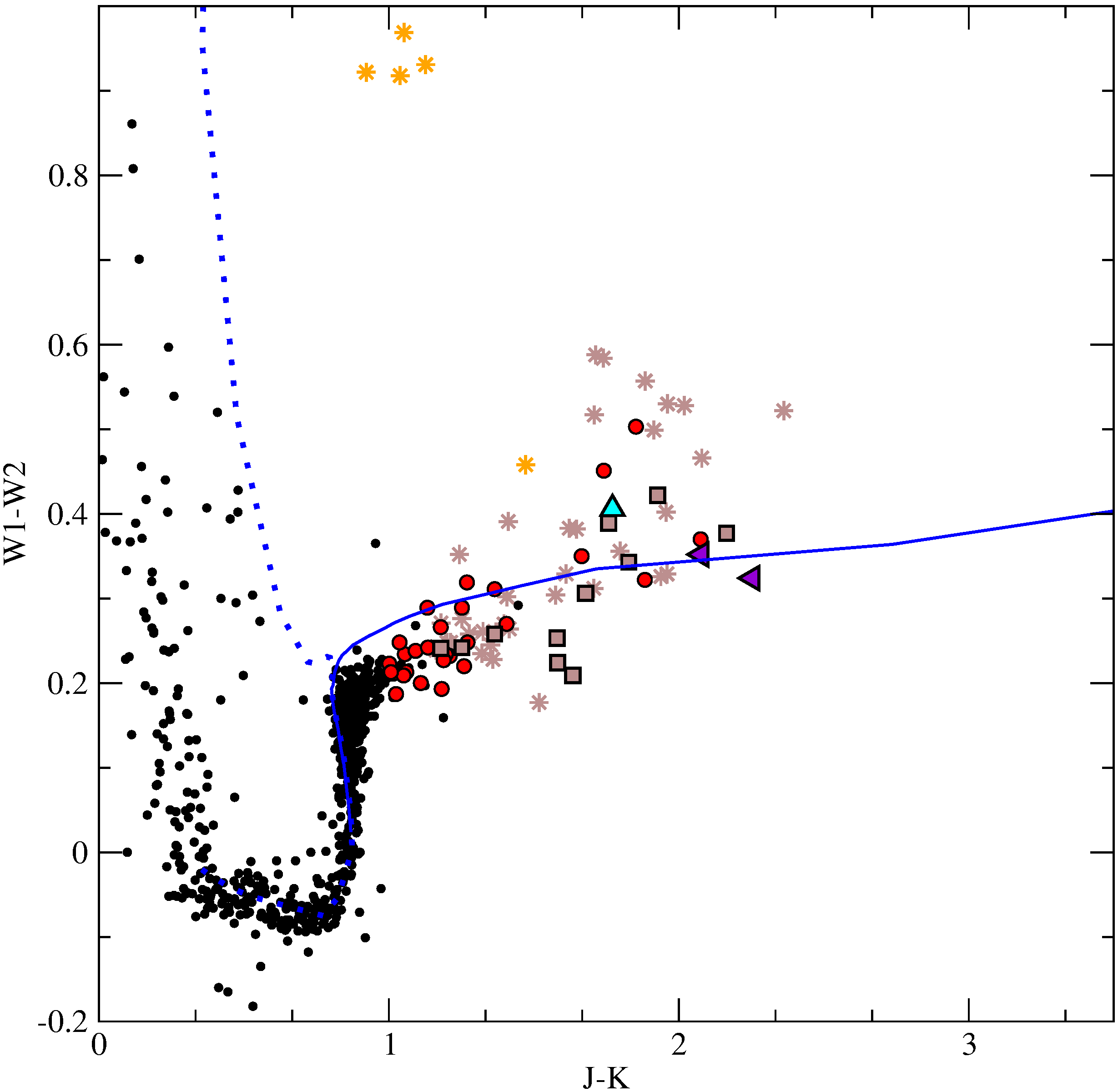}
\caption{($J-K$,$W1-W2$) colour-colour diagram. Symbols are the same as  in Fig.\ 
\ref{fig_Hyades:MJ-J-W1}. Two isochrones for 600 Myr are also plotted: the solid line corresponds to the DUSTY model and the dotted line to the COND model.
The yellow asterisk with the bluest $W1-W2$ colour corresponds to a T0 field dwarf. 
}
\label{fig_Hyades:J-K-W1-W2}
\end{figure}

\section{Contamination}

To assess the number of brown dwarf contaminants included in our final list, we used the surface density, $\rho$, of substellar field objects calculated by \cite{burgasser07}.  Assuming  a mass function $\Psi\left(M\right)=dN/dM\propto M^{-1.5}$ , we obtain that 11 field dwarfs should be found with spectral types between L0 and L3, while 4 and 1 field objects are expected in spectral ranges L3-L6 and L6-T0, respectively.
The assumed mass function is one of the most unfavourable cases, giving on average a larger number of field dwarfs per unit area than other mass functions.
The probability that any of these field objects coincidentally have a proper motion vector pointing to the cluster CP, that is, $\left|\theta_{\rm PM}-\theta\right|\leq 33^\circ$
(see section \ref{Hyades:new_memb_Xmatch}), is about 20\%. Thus, the contamination coming from field dwarfs has an upper limit of two objects within the spectral range L0-L3, one contaminant for L3-L6, a no contamination at all for the L6-T0 range. Candidates in the latter spectral range therefore have a very high probability of being Hyades members.

\section{Spectroscopy}

We obtained near-infrared spectra for three Hyades L dwarf candidates with the
Long-slit Intermediate Resolution Infrared Spectrograph \citep[LIRIS;][]{manchado98,manchado00,manchado04}
on the William Herschel telescope in the Roque de los Muchachos in La Palma, Canary Islands (Spain).
The sky was clear, dark with no moon, and the seeing was in the 0.5$^{\prime\prime}$--1$^{\prime\prime}$ range during our observations.
We used a slit of 1 arcsec and the $zJ$ grism, which covers the 0.89--1.51 micron wavelength range at a
spectral resolution of about 600\@.

The Hyades candidates were visible at the end of the night. We observed 2M0424p0637 and 2M0424p0637 (Table \ref{tab_Hyades:Candidates}) on 8 September 2018
with three and four AB patterns around $UT\,=\,4h30$ and $UT\,=\,3h$, respectively. We set the single on-source
exposures to 300s and an offset of 47 pixels between the A and B positions set to parallactic angle.
We collected four AB patterns for 2M0438p0700 on 9 September 2018 at UT\,=\,3h with on-source integrations
of 360s. We observed a telluric standard of B3 type after each target with an ABBA pattern and on-source
integrations of 1s to 5s to correct for the telluric absorption.
We took dome flats and arc lamps of Xe$+$Ar during the afternoon preceding the observations.

\begin{figure}
  \includegraphics[width=\linewidth, angle=0]{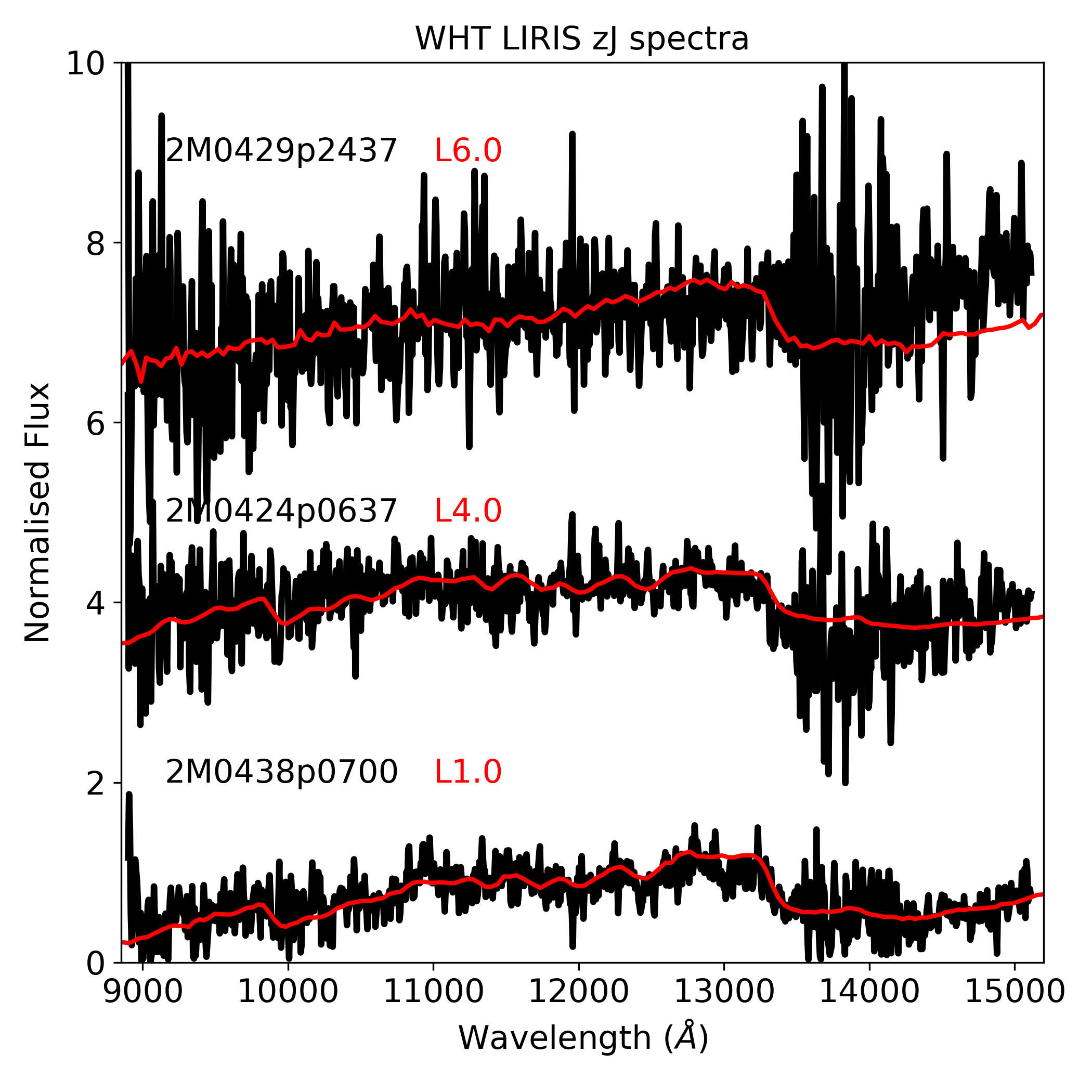}
\caption{Near-infrared spectra for three Hyades L dwarf candidates.
}
\label{fig_Hyades:spec_LIRIS_Hyades_dL}
\end{figure}

We used a master flat field created from a median average of 20 individual frames.
We subtracted the combined dome flat from each of the science frame. We subtracted each
AB pair from each other to produce A$-$B and B$-$A frames. We averaged all A$-$B and B$-$A
frames together and then shifted the combined B$-$A frame to the mean A$-$B frame to
generate a final 2D spectrum of the the target. We optimally extracted the trace of
the target by manually choosing the aperture and background regions. We applied the
wavelength calibration with the Xe$+$Ar arc lamps with an rms better than 0.6\AA{}.
We repeated the process with the telluric standard star. We divided the science
spectrum by the spectrum of the telluric and multiplied by a blackbody of a B3 star
The final spectra, normalised  at 1.24--1.26 microns, are displayed in
Fig.\ \ref{fig_Hyades:spec_LIRIS_Hyades_dL}.
We overplot a few SpeX spectra (\cite{rayner03}) of L dwarfs that we downloaded from the NASA InfraRed Telescope Facility  website and
selected them as templates in the optical. The spectroscopy of all candidates we observed is 
compatible with them being L-type brown dwarfs. By comparing this with low-resolution SpeX spectra, we classify 2M0438$+$0700 as  L4.0$\pm$0.5, 2M0424$+$0637 as  L1$\pm$1 (much earlier than expected; it might be a field object), and 2M0429$+$2437 as L6$-$L8 (hard to classify since L dwarfs from SpeX have lower flux for wavelengths larger than 14500\AA).

\section{Mass function}

\begin{figure}
  \includegraphics[width=\linewidth, angle=0]{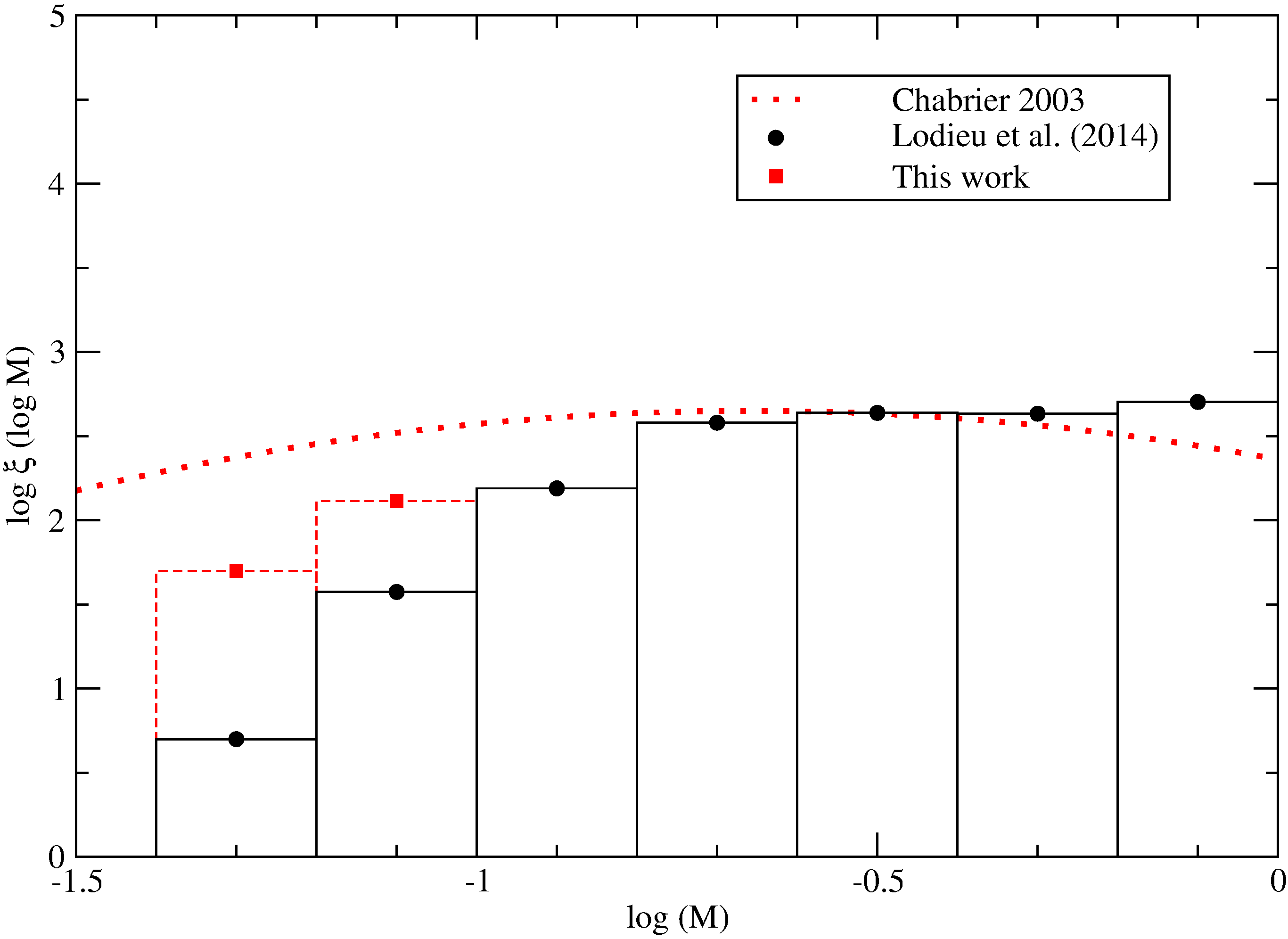}
\caption{ Mass function for the Hyades cluster using data from \cite{lodieu14b} and this work. The log-normal field mass function from \cite{chabrier03} is also overplotted, normalised to one and multiplied by the most populated bin: $\log (M/M_\odot)= -0.5$.}
\label{fig_Hyades:IMF}
\end{figure}

The study of the evolution of the cluster mass function (MF), from the initial to the present mass function, is a powerful tool for investigating the mass segregation of the cluster over time, which is more critical for objects of lowest masses. 
To be able to plot the mass function, $\xi(\log M)=\mathrm{d} N/\mathrm{d}\log M$, the mass of each object from our list must be estimated.
We performed this estimation using the $J-K$ colour from the 2MASS catalogue and the isochrones of the DUSTY model for an age of 700 Myr.
\cite{lodieu14b} computed the mass function for the Hyades cluster using a region with an area similar to the area covered by \cite{hogan08}, which is 275 deg$^2$.  Their mass function is incomplete for objects with masses below the stellar-substellar boundary. We cannot claim that our list of candidates is complete and free of contaminants to obtain definitive conclusions about the Hyades MF. We plot the MF (red dashed line) along with the mass function from \cite{lodieu14b} (black solid line) and the log-normal function of field stars from \cite{chabrier03} (red dotted line) in Fig.\ \ref{fig_Hyades:IMF}. The field mass function has been scaled to the most populated bin: $\log(M/M_\odot)=-0.5$ or $M=0.3M_\odot$. The bin centred 
around $\log \left( M/M_\odot\right)=-1.3$, that is, $M=0.05M_\odot$, must be taken as a lower limit: the coolest objects in this bin are too faint to be detected by the 2MASS survey. By comparing the MF with the log-normal curve from \cite{chabrier03}, we obtain 
a mass loss of $\sim$60\% and $\sim$80\% after scaling for 0.08M$_\odot$ and 0.05M$_\odot$ bins, respectively. These numbers are in good agreement with predictions from numerical simulations
by \cite{adams01}, who found that a 650 Myr old cluster might have lost 70--90\% of its initial substellar population.
We plan to extend the study of the Hyades MF to lower masses making use of deeper surveys such as the UKIRT Infrared Deep Sky Survey Galactic Cluster Survey (\cite{lawrence07}).

%
%
\section{Conclusions}
\label{Hyades:conclusions}

We cross-matched the 2MASS and WISE databases to identify very low mass stars and brown dwarfs with proper motions and photometry consistent with membership to the Hyades cluster. We used Gaia data and a maximum likelihood method to select good Hyades member candidates based on distance considerations. We obtained a sample of 37 objects with proper motions, photometry, and distance (measured by Gaia or estimated with our method) compatible with Hyades membership. 
Sixteen candidates are known from previous studies, and 21 are new. The photometry of 9 of the new candidates places them in the substellar regime.
Two objects show near-infrared colours that place them at the L-T boundary. One of them has an infrared spectrum that indicates an early L type, which means that it could be a background field object, while the other has a spectrum that is compatible with L6$-$L8.
If confirmed spectroscopically, these two objects would be the latest L dwarf members in the Hyades and would bridge the gap between mid-L dwarfs and the two T dwarfs that have previously been reported \citep{bouvier08a}.  We have also calculated the mass function for the mass range of our candidates. 
%

%
%
\begin{acknowledgements}
This research has been supported by the Spanish Ministry of 
Economy and Competitiveness (MINECO) under the grants AYA2015-69350-C3-2-P and 
AYA2015-69350-C3-3-P.



 This article is based on observations made in the Observatorios de Canarias del IAC with the WHT operated on the island of La Palma by the Isaac Newton Group of Telescopes in the Observatorio del Roque de los Muchachos.

This publication makes use of data products from the Two Micron All Sky Survey
(2MASS), which is a joint project of the University of Massachusetts and
the Infrared Processing and Analysis Center/California Institute of Technology,
funded by the National Aeronautics and Space Administration and the National
Science Foundation. 
This publication makes use of data products from the Wide-field Infrared Survey Explorer, which is a joint project of the University of
California, Los Angeles, and the Jet Propulsion Laboratory/California Institute
of Technology, funded by the National Aeronautics and Space Administration.


Funding for the Sloan Digital Sky Survey IV has been provided by the Alfred P. Sloan Foundation, the U.S. Department of Energy Office of Science, and the Participating Institutions. SDSS acknowledges support and resources from the Center for High-Performance Computing at the University of Utah. The SDSS web site is www.sdss.org.

SDSS-IV is managed by the Astrophysical Research Consortium for the 
Participating Institutions of the SDSS Collaboration including the 
Brazilian Participation Group, the Carnegie Institution for Science, 
Carnegie Mellon University, the Chilean Participation Group, the French Participation Group, Harvard-Smithsonian Center for Astrophysics, 
Instituto de Astrof\'isica de Canarias, The Johns Hopkins University, 
Kavli Institute for the Physics and Mathematics of the Universe (IPMU) / 
University of Tokyo, Lawrence Berkeley National Laboratory, 
Leibniz Institut f\"ur Astrophysik Potsdam (AIP),  
Max-Planck-Institut f\"ur Astronomie (MPIA Heidelberg), 
Max-Planck-Institut f\"ur Astrophysik (MPA Garching), 
Max-Planck-Institut f\"ur Extraterrestrische Physik (MPE), 
National Astronomical Observatory of China, New Mexico State University, 
New York University, University of Notre Dame, 
Observat\'ario Nacional / MCTI, The Ohio State University, 
Pennsylvania State University, Shanghai Astronomical Observatory, 
United Kingdom Participation Group,
Universidad Nacional Aut\'onoma de M\'exico, University of Arizona, 
University of Colorado Boulder, University of Oxford, University of Portsmouth, 
University of Utah, University of Virginia, University of Washington, University of Wisconsin, 
Vanderbilt University, and Yale University.

This research has made use of the Simbad and Vizier databases, operated
at the Centre de Donn\'ees Astronomiques de Strasbourg (CDS), and
of NASA's Astrophysics Data System Bibliographic Services (ADS) and
 has made use of the NASA/ IPAC Infrared Science Archive, which is operated by the Jet Propulsion Laboratory, California Institute of Technology, under contract with the National Aeronautics and Space Administration."
This work has made use of data from the European Space Agency (ESA) mission
{\it Gaia} (\url{https://www.cosmos.esa.int/gaia}), processed by the {\it Gaia}
Data Processing and Analysis Consortium (DPAC,
\url{https://www.cosmos.esa.int/web/gaia/dpac/consortium}). Funding for the DPAC
has been provided by national institutions, in particular the institutions
participating in the {\it Gaia} Multilateral Agreement.

\end{acknowledgements}
%

%
%
\bibliographystyle{aa}
\bibliography{biblio} 

%

\newpage

\clearpage

\begin{table}
\caption{
List of the 36 best Hyades candidates after applying all selection criteria described in this work along with their 
photometric data. }
\label{tab_Hyades:Candidates}
\begin{tabular}{l l l l l l l l l l}
\hline
Object   &   Ra &    Dec  & $i$ & $J$ &  $W2$ & $i-z$ &$J-K_s$&  $W1-W2$ &  Other Name\\
\hline
2M0353$+$1030&03:53:08.45&$+$10:30:56.5&19.45&15.45&13.46&1.88&1.27&0.31&...\\
2M0355$+$1439&03:55:20.15&$+$14:39:29.7&17.77&13.83&12.12&1.71&1.13&0.29&...\\
2M0408$+$0742&04:08:10.32&$+$07:42:49.5&13.59&11.91&...&1.17&0.24 & ...& ... \\
2M0409$+$1247&04:09:02.72&$+$12:47:29.2&...&16.42&13.75&...&1.89&0.32&...\\
2M0410$+$1459&04:10:23.91&$+$14:59:10.4&19.94&15.75&13.68&1.77&1.58&0.25& Hya03 $^a$\\
2M0416$+$2052&04:16:56.51&$+$20:52:36.4&17.10&13.66&12.20&1.46&1.05&0.23&...\\
2M0418$+$2131&04:18:34.83&$+$21:31:27.5&22.77&17.15&13.89&2.43&1.77&0.41& 2M0418$+$21$^b$\\
2M0420$+$2356&04:20:24.42&$+$23:56:13.4&...&14.60&12.85&...&1.18&0.24&Hya01$^a$\\
2M0420$+$1345&04:20:50.17&$+$13:45:53.1&...&14.27&12.63&...&1.21&0.23&LH0418$+$13\\
2M0421$+$2023&04:21:45.87&$+$20:23:44.6&17.31&13.94&12.48&1.49&1.01&0.22&...\\
2M0422$+$1358&04:22:05.12&$+$13:58:47.4&...&15.50&13.64&...&1.25&0.29&Hya06$^a$\\
2M0424$+$0637&04:24:18.57&$+$06:37:44.9&22.32&17.15&14.20&2.74&1.85&0.5&...\\
2M0429$+$2437&04:29:30.24&$+$24:37:49.7&...&17.50&14.46&...&2.07&0.37&...\\
2M0429$+$2529&04:29:47.24&$+$25:29:18.9&17.70&14.12&12.58&1.57&1.1&0.2&...\\
2M0433$+$1611&04:33:50.93&$+$16:11:03.2&...&17.26&14.31&...&1.74&0.45&...\\
2M0435$+$2008&04:35:13.55&$+$20:08:01.4&17.14&13.70&12.08&1.45&1.09&0.24&...\\
2M0435$+$1927&04:35:20.24&$+$19:27:47.3&19.71&15.43&13.65&1.84&1.26&0.22&...\\
2M0435$+$1215&04:35:51.76&$+$12:15:20.0&...&13.68&12.23&...&1&0.21&...\\
2M0436$+$1151&04:36:27.67&$+$11:51:24.4&...&13.87&12.12&...&1.19&0.24&...\\
2M0436$+$1901&04:36:42.68&$+$19:01:35.2&21.70&17.12&13.87&1.91&2.25&0.32&WISEAJ043642.75$+$190134.8$^c$\\
2M0438$+$0700&04:38:03.50&$+$07:00:55.1&...&16.91&14.07&...&1.66&0.35&...\\
2M0440$+$2325&04:40:10.99&$+$23:25:14.2&...&13.38&11.98&...&1&0.22&...\\
2M0441$+$2130&04:41:05.48&$+$21:30:02.0&21.70&17.27&14.20&1.38&2.07&0.35&WISEAJ044105.56$+$213001.5$^c$\\
2M0441$+$1453&04:41:45.15&$+$14:53:58.3&...&14.69&12.64&...&1.41&0.27&...\\
2M0444$+$1901&04:44:23.26&$+$19:01:37.8&...&15.49&13.75&...&1.18&0.19&...\\
2M0445$+$1443&04:45:13.14&$+$14:43:26.8&...&14.30&12.77&...&1.14&0.25&...\\
2M0445$+$1503&04:45:33.05&$+$15:03:02.6&...&14.03&12.52&...&1.06&0.21&...\\
2M0445$+$1246&04:45:43.71&$+$12:46:31.6&18.90&15.14&13.13&1.68&1.27&0.25&...\\
2M0446$+$1857&04:46:09.64&$+$18:57:28.1&...&13.77&12.36&...&1.05&0.21&...\\
2M0447$+$1719&04:47:14.95&$+$17:19:49.7&...&13.44&12.03&...&1.01&0.21&...\\
2M0448$+$2051&04:48:22.45&$+$20:51:43.3&16.64&13.30&11.84&1.42&1.02&0.19&...\\
2M0453$+$2033&04:53:34.26&$+$20:33:51.7&...&15.32&12.95&...&1.37&0.31&...\\
2M0455$+$2140&04:55:58.98&$+$21:40:00.8&...&14.11&12.39&...&1.18&0.27&...\\
2M0458$+$1212&04:58:45.66&$+$12:12:34.3&19.61&15.60&13.49&1.73&1.58&0.23&Hya08$^a$\\
2M0459$+$1304&04:59:32.54&$+$13:04:54.9&18.63&14.63&12.92&1.82&1.19&0.23&...\\
2M0500$+$1207&05:00:19.36&$+$12:07:34.2&...&13.94&12.49&...&1.04&0.25&...\\
\hline
\hline
\end{tabular}
\tablefoot{
\tablefoottext{a}{\citet{hogan08}}
\tablefoottext{b}{\citet{perezgarrido17, lodieu18}}
 \tablefoottext{c}{\citet{schneider17}}
}
\end{table}

\newpage
\begin{table*}
\tiny
\caption{Proper motions of candidates estimated in our cross-match (Cols. 2 and 3) and from Gaia DR2 (Cols. 4 and 5). Column 6 show the estimated distance using the maximum likelihood method developed for this work. Errors in Col. 6 are the dispersion of distance estimates carried out in the five bands considered in the maximum likelihood method, see text. Columns 7 and 8 shows distances measured by the Gaia mission and by other authors, respectively. Estimated spectral types with our method are listed in Col. 9, while types reported by other authors are listed in the last column.}
\begin{tabular}{l r r r r c c c l l }
\hline
Object   & $\mu_\alpha$ & $\mu_\delta$&  $\mu^{\rm Gaia}_\alpha$ & $\mu^{\rm Gaia}_\delta$& d$^{\rm Phot.} $ 
& d$^{\rm Gaia}$ & d$^{\rm others}$& SpType & SpType \\
         &   (mas/yr)     &    (mas/yr)    & (mas/yr)                 & (mas/yr)                 & (pc) 
&  pc & pc &(Estimated)& (Reported)\\
\hline
2M0353$+$1030   &    133    &     $-$18    &      127  &   $-$22  &      52$\pm$  1.3  &     43.1$^{+2.2}_{-2.0}$  &  ...    &    L0.7    &  L0.9V$^1$ \\
2M0355$+$1439   &    148    &     $-$16    &      154  &   $-$20  &      29$\pm$  0.4  &     36.8$^{+0.3}_{-0.3}$  &  ...    &    M9.8    &  M8.8V$^1$ \\
2M0408$+$0742     &    145    &        84   &       142 & 73        &      28$\pm$ 0.2 & 35.7$^{+0.3}_{-0.3}$ &...  &        M9.2 & M8.1V$^1$ \\
2M0409$+$1247   &    106    &     $-$10   &  ...      & ...  &      39$\pm$  2.1         &      ...      &    ...      &    L4.5    &... \\
2M0410$+$1459   &    107    &     $-$22   &   108  & $-$11 &     64$\pm$3.8              &  57.7$^{+5.8}_{-4.8}$ & 62.1$^{+4.7}_{-4.9}$ $^2$      & M9.6 &L0.5$^2$ \\
2M0416$+$2052   &    112    &     $-$31   &     119  &   $-$39  &      40$\pm$  0.5  &      46.6$^{+0.6}_{-0.6}$ &...&   M7.1    & ...   \\
2M0418$+$2131   &    120    &     $-$55   &... & ...   &      37$\pm$  2.9  &...&48.8$\pm$4.0$^a$&    L7.1    &   L5$^3$ \\
2M0420$+$2356   &    126    &     $-$36   &     130  &   $-$29  &      39$\pm$  0.3  &      44.2$^{+0.9}_{-0.9}$   & 48.9$^{+3.8}_{-3.4}$ $^2$   &    L0.1    &  M8.5$^2$ \\
2M0420$+$1345   &    109    &     $-$9    &    116  &   $-$21  &      40$\pm$  0.5  &      42.5$^{+1.0}_{-0.9}$    & ...   &    M8.9    &  M8.5 \\
2M0421$+$2023   &    96     &     $-$36   &  104  &   $-$33  &      47$\pm$  0.1  &      49.1$^{+0.6}_{-0.6}$      & ... &  M6.9    & ...    \\
2M0422$+$1358   &    84     &     $-$10   & 89   &  $-$18  &      66$\pm$  1.9  &54.9$^{+2.9}_{-2.6}$   &64.3$^{+4.5}_{-4.7}$ $^c$  &   M9.1    &  M9.5$^2$ \\
2M0424$+$0637   &    159    &     65    & ...  &  ...    &      31$\pm$  2.0  &... &... &   L9.1    &  f L1$\pm$1.0$^4$ \\
2M0429$+$2437   &    45     &     $-$71   &...  &... &      42$\pm$  1.9  &... &...  &  L7.2    &  L6.0$-$L8.0$^4$  \\
2M0429$+$2529   &    93     &     $-$53   &98  &   $-$52  &      49$\pm$  0.7  &      48.0$^{+0.7}_{-0.7}$ & ...     &    M7.1    &  M8V \\
2M0433$+$1611   &    73     &     $-$24   & ...  &...&      38$\pm$  1.4 &... & ...&    L8.0    & ...  \\
2M0435$+$2008   &    91     &     $-$43   &101  &   $-$43    &      27$\pm$  0.8  &      45.6$^{+0.4}_{-0.4}$   & ...   &    L0    & ...  \\
2M0435$+$1927   &    174    &     $-$46   &179  &   $-$39    &      58$\pm$  0.5  &      52.1$^{+2.4}_{-2.2}$   & ...   &    M9.9    & ...  \\
2M0435$+$1215   &    75     &     1     &92  &   $-$13    &      42$\pm$  0.3  &      50.6$^{+0.4}_{-0.4}$ & ...     &    M7.0    & ...  \\
2M0436$+$1151   &    94     &     $-$11   &100  &   $-$8  &      28$\pm$  0.3  &      43.7$^{+0.5}_{-0.5}$ &  ...     &    L0.2    &  M9$^5$ \\
2M0436$+$1901   &    93     &     $-$30   &...    & ...     &      33$\pm$  2.1  & ...  &35$\pm$4$^b$     &    L7.0    &  L5$^6$  \\
2M0438$+$0700   &    82     &     23    & ...   &  ...    &      44$\pm$  1.0  &... & ... &  L5.1    &  L4.0$\pm$0.5$^4$  \\
2M0440$+$2325   &    85     &     $-$29   &90  &   $-$46   &      38$\pm$  0.6  &      50.1$^{+0.4}_{-0.4}$ &   ...    &    M6.8    & ...   \\
2M0441$+$2130   &    96     &     $-$31   & ...  &...&      38$\pm$  1.4  & ... & 45$\pm$9$^b$      &   L7.4    &  L6$^6$  \\
2M0441$+$1453   &    87     &     $-$15   &98  &   $-$23   &      37$\pm$  1.6  &      45.8$^{+1.3}_{-1.2}$ &  ...     &    L0.4    & ...   \\
2M0444$+$1901   &     79    &     $-$30   &  74 &  $-$32   &    61$\pm$0.9  &                61.7$^{+4.0}_{-3.5}$&...  &    M9.9  & ... \\
2M0445$+$1443   &    80     &     $-$15   &82  &   $-$22   &      49$\pm$  0.8  &      50.8$^{+1.0}_{-0.9}$     & ...  &    M7.6    & ...   \\
2M0445$+$1503   &    84     &     $-$8    &84  &   $-$23   &      49$\pm$  0.6  &      48.1$^{+0.5}_{-0.5}$     & ...  &    M6.9    & ...   \\
2M0445$+$1246   &    88     &     $-$12   &100  &   $-$16   &      40$\pm$  0.5  &      43.4$^{+1.5}_{-1.4}$     &...   &    L1.5    & ...   \\
2M0446$+$1857   &    82     &     $-$36   &86  &   $-$38   &      44$\pm$  0.9  &      48.8$^{+0.5}_{-0.5}$     & ...  &    M7.0    & ...   \\
2M0447$+$1719   &    85     &     $-$21   &95  &   $-$31   &      38$\pm$  0.4  &      43.9$^{+0.3}_{-0.3}$    & ...   &    M6.8    & ...   \\
2M0448$+$2051   &    67     &     $-$27   &76  &   $-$37  &      35$\pm$  0.3  &      55.9 $^{+0.8}_{-0.7}$   & ...   &    M6.9    &  M6V$^7$ \\
2M0453$+$2033   &    58     &     $-$90   & 68 & $-$89   &      31$\pm$0.4  & 33.8$^{+1.8}_{-1.6}$ &...& L3.5   & ...\\
2M0455$+$2140   &    71     &     $-$39   &77  &   $-$45 &      37$\pm$  0.4  &      48.0 $^{+0.8}_{-0.7}$ &  ...    &    M8.9    &  M8.5V$^7$ \\
2M0458$+$1212   &    82     &     $-$12   &86  &   $-$16  &      50$\pm$  2.8  &      41.2  $^{+1.7}_{-1.6}$ & 57.9$^{+4.4}_{-4.6}$ $^6$     &    L0.8    &    L0.5$^2$ \\
2M0459$+$1304   &    71     &     $-$22   &80  &   $-$21   &      42$\pm$  0.3  &      48.2 $^{+1.2}_{-1.1}$  & ...    &    M9.7    &   ... \\
2M0500$+$1207   &    104    &     $-$7    &98  &   $-$17   &      46$\pm$  0.6  &      48.2 $^{+0.5}_{-0.5}$  & ...    &    M7.2    &  ...  \\

\hline
\end{tabular}

References:
(1) {\citet{bardalez14}}; (2) {\citet{lodieu14b}}; (3) {\citet{perezgarrido17, lodieu18}};
(4) {This work}; (5) {\citet{faherty09}}; (6) {\citet{schneider17}};
(7) {\citet{luhman06}}


\label{tab_Hyades:likelihoodResults}
\end{table*}

\end{document}